\documentclass[aps,pra,preprint,showpacs]{revtex4-1}
\usepackage{amsmath,amssymb}
\usepackage[mathscr]{eucal}
\usepackage{color}

\def\DJo{$\;$\kern-.4em \hbox{D\kern-.8em\raise.15ex\hbox{--}\kern.35em okovi\'c}}
\def\CC{{\rm\kern.24em \vrule width.04em height1.46ex depth-.07ex
\kern-.30em C}}
\def\RR{{\rm
         \vrule width.04em height1.58ex depth-.0ex
         \kern-.04em R}}
\def\P{{\rm I\kern-.25em P}}
\def\id{{\rm 1\kern-.22em l}}

\newcommand{\ket}[1]{\left | #1 \right\rangle}

\begin{document}

\title{Polynomial invariants for discrimination and classification
      of four-qubit entanglement 
      }
\author{Oliver Viehmann}
\affiliation{Physics Department,
             Arnold Sommerfeld Center for Theoretical Physics,
             and Center for NanoScience,\\
             Ludwig-Maximilians-Universit\"at,
             Theresienstra{\ss}e 37,
             D-80333 M\"unchen, Germany}
\author{Christopher Eltschka}
\affiliation{Institut f\"ur Theoretische Physik, 
         Universit\"at Regensburg, D-93040 Regensburg, Germany}
\author{Jens Siewert}
\affiliation{Departamento de Qu\'{\i}mica F\'{\i}sica, Universidad del Pa\'{\i}s Vasco -
             Euskal Herriko Unibertsitatea, Apdo. 644, 48080 Bilbao, Spain}
\affiliation{Ikerbasque, Basque Foundation for Science, Alameda Urquijo 36, 48011 Bilbao, Spain}

\begin{abstract}
It is well known that the number of entanglement classes in SLOCC
(stochastic local operations and classical communication) 
classifications increases with the number of qubits and 
is already infinite for four qubits~\cite{Duer2000}. Bearing in 
mind the rapid evolution of experimental technology, criteria for 
explicitly discriminating and classifying pure states of four and 
more qubits are highly desirable and therefore in the focus of 
intense theoretical 
research~\cite{Sudbery1999,Verstraete2002,Miyake2004,OS2005,Lamata2006,Mandilara2006,Lamata2007,Cao2007,Chterental2007,Bastin2009,Duff2010,Li2009,Kraus2010,Winter2010}.
In this article
we develop a general criterion for the discrimination of
pure $N-$partite entangled states in terms of polynomial
$SL(d,\CC)^{\otimes N}$ invariants. 
By means of this criterion, existing
SLOCC classifications of four-qubit entanglement are reproduced.
Based on this we propose a polynomial
classification scheme in which families are identified through
``tangle patterns'', thus bringing together qualitative and 
quantitative description of entanglement.
\end{abstract}
\maketitle

Polynomial functions of the coefficients of pure quantum states play
an important role as entanglement measures~\cite{Wootters1997,Duer2000,CKW2000,Wong2001,Verstraete2003,Luque2003,Leifer2004,OS2005,Gour2010}. Such measures have to be
homogeneous and
invariant under local $SL$ transformations~\cite{Duer2000,Verstraete2003}.
For pure two-qubit and three-qubit states,
concurrence~\cite{Wootters1997} and three-tangle~\cite{CKW2000} are
the unique polynomials of this kind~\cite{Brylinski2002}. 
An important step towards their generalization has been taken
by Luque and Thibon~\cite{Luque2003}
who found the generators of the algebra of ${SL}(2,\CC)^{\otimes 4}$-invariant 
polynomials in the coefficients of pure four-qubit states. 
Further, in Refs.~\cite{Leifer2004,OS2005} general methods to construct 
$SL(2,\CC)^{\otimes N}$-invariant polynomials for $N$-qubit states were introduced.
However, even for four qubits it is not yet clear, whether there is a choice of
polynomials that properly generalizes concurrence and three-tangle. 

Presumably the most interesting consequence deriving from the properties of invariant 
polynomials in the cases of two and three qubits is that they impose the 
\textit{SLOCC classification} of entangled states~\cite{Duer2000}.
There is only one type of entanglement for two qubits
and the concurrence is non-vanishing exactly for the entangled states.
In multipartite systems, 
there may occur different types of entanglement.
For example pure three-qubit entangled states
may be GHZ-type states (non-vanishing three-tangle), or 
$W$-type states (vanishing three-tangle)~\cite{Duer2000}.
That is, concurrence and three-tangle quantify {\em class-specific} 
entanglement.
For four and more qubits
the number of SLOCC classes is infinite~\cite{Duer2000}. 
Therefore, the general idea of any SLOCC classification, e.g.,
Refs.~\cite{Verstraete2002,Miyake2004,Lamata2007,Cao2007,Chterental2007,Bastin2009,Duff2010},
is to arrange the representatives
of the infinitely many classes
into a finite number of sets according to some  SLOCC-invariant criterion, 
such as the {\em Schmidt measure}~\cite{Duer2000,EisertBriegel2001}, the {\em degeneracy configuration}~\cite{Bastin2009} 
or the structure of the right singular subspace of the state coefficient 
matrix~\cite{Lamata2007}. 
Obviously, all classifications comprise exactly the same classes -- merely the arrangement into
sets (which, henceforth, we call {\em families}) is different. The families
are defined by the representative states they contain, thus providing a coarse graining to the SLOCC classes. At least
one family comprises an infinite number of them. 

In the existing classifications of pure four-qubit states 
it is not easy to determine 
to which class or family a given
arbitrary state belongs, 
nor are they readily generalized to more complicated Hilbert spaces.
Whether there 
exists a general relation between polynomial invariants and SLOCC 
classification analogous to the cases of two  and three qubits is also still unknown.
The only efforts touching upon this question
were made in Refs.~\cite{Chterental2007,Cao2007,Li2009},
however without a compelling answer. 
The aim of this article is to fill these gaps  and 
to bring qualitative and quantitative 
aspects of entanglement theory in line. To begin with, we introduce a strong  sufficient
criterion  for distinguishing SLOCC classes of arbitrary multipartite states 
based on polynomial invariants.  
We illustrate its application by  extending
earlier findings by Li {\em et al.}~\cite{Li2009}.
Subsequently,  we show that the classifications of symmetric 
and of general four-qubit states due to Bastin {\em et al.}~\cite{Bastin2009} and 
Lamata {\em et al.}~\cite{Lamata2007}, respectively,
can be almost completely matched with certain sets of four-qubit polynomial invariants.
As our main result, we then propose a polynomial-based SLOCC classification.

{\em Polynomial discrimination criterion for pure multipartite states. -- }
Two pure quantum states $\psi^{(N)}$ and $\bar{\psi}^{(N)}$ of an $N$-partite Hilbert space 
$\mathcal{H}^{(N)}=\mathcal{H}_1 \otimes \ldots \otimes \mathcal{H}_N$ are 
interconvertible 
with a nonzero chance by means of SLOCC if and only if there exist invertible operators 
$J_1,\ldots J_N$ acting on the local Hilbert spaces $\mathcal{H}_j$ (of dimension $d_j$)
such that \cite{Duer2000}
\begin{align}\label{SLOCC}
|\bar{\psi}^{(N)}\rangle\  =\ J_1 \otimes \ldots \otimes J_N \, |\psi^{(N)}\rangle \ \ .
\end{align}
Throughout this paper we consider unnormalized vectors $\psi^{(N)}$. 
By means of the definition in equation~\eqref{SLOCC} SLOCC interconvert\-ibility 
imposes an equivalence relation on the set of all vectors of $\mathcal{H}^{(N)}$. 
The SLOCC equivalence classes are sets of states with equivalent 
multipartite entanglement in the sense that, under SLOCC,
the same tasks can be performed with them.

Suppose $\mathcal{P}_{[i]}$ and $\mathcal{P}^{\prime}_{[j]}$ are
homogeneous  functions of degrees $i$ and $j$ of the states in a
Hilbert space $\mathcal{H}^{(N)}$ that are invariant
under $SL(d_k,\CC)^{\otimes N}$ transformations 
where $d_k = \dim{\{\mathcal{H}_k\} }$.
Then, for integers $m,n$ 
with $im=jn$,  and a fixed state $\psi^{(N)}$,  a complex number $\eta$
exists such that
\begin{align}\label{eq: defalpha}
\big( \mathcal{P}_{[i]}(\psi^{(N)}) \big) ^m 
\ = \ 
\eta \, \cdot \big( \mathcal{P}^{\prime}_{[j]}(\psi^{(N)}) \big) ^n \ \ .
\end{align}
Here $\eta$ is unique and invariant under SLOCC transformations on $\psi^{(N)}$ 
as long as $\mathcal{P}^{\prime}_{[j]}$ is different 
from zero. That is, for
$\bar{\psi}^{(N)}\ =\ J_1 \otimes \ldots \otimes J_N \, \psi^{(N)}$
 we have also
$
\left( \mathcal{P}_{[i]}(\bar{\psi}^{(N)}) \right) ^m 
\ = \ 
\eta \  \cdot \big( \mathcal{P}^{\prime}_{[j]}(\bar{\psi}^{(N)}) \big) ^n$.
The ratio of homogeneous $SL(d_k,\CC)^{\otimes N}$ invariants of the same degree 
is invariant under SLOCC. Therefore,
{\em  for two SLOCC-equivalent states the 
     ratio of two arbitrary invariants must 
     be the same}. The spirit of 
this criterion has been applied before~\cite{Luque2003,Li2009}, however without emphasizing its generality.

An important consequence 
is that from two independent invariants (which for more than three qubits
can always be found~\cite{OS2005,Dokovic2009})
one can construct an invariant that 
vanishes for a given SLOCC class due to
\begin{align}\label{eq: zero for a class}
\mathcal{P}_{[i]} ^m - \eta \, \cdot \mathcal{P}^{\prime \, n}_{[j]}\  =\  0\ \ .
\end{align}

In the following, we will focus exclusively on  polynomial invariants and 
the four-qubit Hilbert space, 
since in that case all generators of the algebra of polynomial 
invariants~\cite{Luque2003,Dokovic2009} are known. 
Following the notation of Ref.~\cite{Dokovic2009} we define 
\begin{align}
(( A_1 \bullet \ldots \bullet A_N )) &\ =\ 
     \left\langle \psi^{\ast} |  A_1  \psi \right\rangle \cdot \ldots \cdot 
     \left\langle \psi^{\ast} | A_N \psi \right\rangle \ \  
\\
\sigma_\mu \bullet \sigma^\mu & \ =\  
    \sum_{\mu=0}^3{g_\mu \cdot \sigma_\mu \bullet \sigma_\mu} 
\end{align}
for operators $A_i$ that act on the Hilbert space of $\psi$, 
the Pauli matrices 
$(\sigma_0,\sigma_1,\sigma_2,\sigma_3) = (\id_2,\sigma_x,\sigma_y,\sigma_z)$ and 
$(g_0,g_1,g_2,g_3):=(-1,1,0,1)$. The $\bullet$ symbol denotes
a tensor product that refers to copies of the same state whereas
we do not specify tensor products between the parties:
$ \ldots \sigma_\mu \sigma_\nu \ldots  \equiv  
                  \ldots \sigma_\mu \otimes \sigma_\nu \ldots 
$
As generators for the $SL(2,\CC)^{\otimes 4}$-invariant polynomials 
we may choose, e.g., 
\begin{subequations}\label{def:gen}
\begin{align}
\mathcal A_{[2]}&= ((\sigma_2 \sigma_2 \sigma_2 \sigma_2))\label{def:A}\\
\mathcal B^{I}_{[4]}&=((\sigma_\mu \sigma_\nu \sigma_2 \sigma_2 \bullet \sigma^\mu \sigma^\nu \sigma_2 \sigma_2))\\
\mathcal B^{II}_{[4]}& =((\sigma_\mu \sigma_2 \sigma_\nu \sigma_2 \bullet \sigma^\mu \sigma_2 \sigma^\nu \sigma_2))\\
\mathcal C_{[6]} &=((\sigma_\mu \sigma_\nu \sigma_2 \sigma_2 \bullet \sigma^\mu \sigma_2 \sigma_\lambda \sigma_2 \bullet \sigma_2 \sigma^\nu \sigma^\lambda \sigma_2))\label{def:C}
\ \ .
\end{align}
\end{subequations}
This set is complete. 
The generator $\mathcal A_{[2]}$ is the well-known four-concurrence~\cite{Wong2001} 
and $\mathcal C_{[6]}$ was introduced in Ref.~\cite{OS2005}. We define
$\mathcal{B}^{III}_{[4]}$ via the sum rule~\cite{Dokovic2009} ${\mathcal B}^I+{\mathcal B}^{II}+{\mathcal B}^{III}=3{\mathcal A}^2$ 
(we omit the subscript indicating the homogeneous degree).  
Note that the polynomials ${\mathcal B}^j$ are
not invariant under qubit permutation.

The knowledge of all generators
allows to exhaustively exploit our criterion 
to distinguish the SLOCC classes of two four-qubit states $\psi^{(4)}$ and $\bar{\psi}^{(4)}$.
We introduce the abbreviations
$\mathcal A(\psi^{(4)})=\alpha ,\ \mathcal A(\bar{\psi}^{(4)})=\bar{\alpha},\
\mathcal B^{I}(\psi^{(4)})=\beta_1,\ldots, 
\mathcal C(\bar{\psi}^{(4)}) =\bar{\gamma}$.
Our criterion leads to the following equations that can be
checked in order to decide whether the states
$\psi^{(4)}$ and $\bar{\psi}^{(4)}$ may belong to the same SLOCC class:
\begin{align}\label{eq:criterion4}
\left( \dfrac{\alpha}{\bar{\alpha}} \right)^2 =  \dfrac{\beta_1}{\bar{\beta}_1} 
                                              =  \dfrac{\beta_2}{\bar{\beta}_2} 
\ \ , \ \ \
\left( \dfrac{\alpha}{\bar{\alpha}} \right)^3 = \dfrac{\gamma}{\bar{\gamma}} \ \ .
\end{align}
If, e.g., $\bar{\alpha}$ is zero, its counterpart $\alpha$ must be zero as well, otherwise
$\psi^{(4)}$ and $\bar{\psi}^{(4)}$ cannot be SLOCC equivalent. In contrast,
if all equations in \eqref{eq:criterion4} hold, the states are not necessarily SLOCC interconvertible.

For example, Li {\em et al.}~\cite{Li2009} 
studied states of the $G_{abcd}$ family from Ref.~\cite{Verstraete2002}
with $b=c=0$ and $a,d\neq 0$.
They concluded that this family can be split into three sub-families,
two of which contain only a single SLOCC class.  
Whether this is also the case 
for the third subfamily (A1.3) remained an open question. 
 We can easily 
negate it by means of our criterion as formulated in equations~\eqref{eq:criterion4}.
The generators in Eq.~(\ref{def:gen})
yield for the corresponding states 
\begin{align}
\begin{split}\label{Li}
\alpha= a^2 + d^2 \qquad \beta_1= 3 a^4 -2 a^2 d^2 + 3 d^4 \\
\beta_2 = 4 a^2 d^2 \qquad \gamma= -4 a^2 d^2 (a^2 + d^2)\ \ .
\end{split}
\end{align}
One sees that equations~\eqref{eq:criterion4} can be violated 
(e.g., $a=d=\bar{a}=1,\, \bar{d}=\sqrt{2}$). 
Consequently, subfamily A1.3 contains more than one SLOCC class.

{\em Polynomial classification of symmetric four-qubit states. -- }
Now we turn to a classification of symmetric four-qubit states $\psi^{(4)}_S$ 
which was presented by Bastin {\em et al.}~\cite{Bastin2009}. 
Five so-called degeneracy configurations ${\mathscr D}_{\{ n_i\}}$ 
define the five families of 
the SLOCC classification, see Table I. 
\begin{table}[htb]
\begin{tabular}{cccccc}
\hline \hline
 ${\mathscr D}_{\{n_i\}}$  & representative & $\mathcal A$  & $\mathcal C$ 
                                                  & $\mathcal D$ & type \\[1mm]
\hline
 ${\mathscr D}_{4}$  &    $  D_4^{(0)}$ &   0   &  0       &    0    & separable  \\
 ${\mathscr D}_{3,1}$  &    $  D_4^{(1)}$ &   0   &  0       &    0    & $W$  \\
 ${\mathscr D}_{2,2}$  &    $   D_4^{(2)}$ &   1   &  -5/9     &    0    & 
                                           $D_4^{(2)}$ \\
 ${\mathscr D}_{2,1,1}$  &  $  D_4^{(0)}+D_4^{(2)}$ &  1  & -5/9 &  0 & 
                                           $D_4^{(2)} $\\
 ${\mathscr D}_{1,1,1,1}$  &  $  \ket{0000}+\ket{1111}+\mu D_4^{(2)}$ &   $a(\mu)$   
                 &  $ c(\mu)$
                 &  $ d(\mu)$
                 & $X$  \\[1mm]
\hline\hline
\end{tabular}
\caption{Comparison of the polynomial characterization and the SLOCC 
         classification of symmetric four-qubit states~\cite{Bastin2009}.
         Note that for symmetric states the sum rule for the generators 
         implies 
         ${\mathcal B}^j={\mathcal A}^2$. 
         The representatives are given in the basis of the symmetric
         four-qubit Dicke states $D_4^{(k)}$ with $k$ $\ket{1}$ components.
         For the continuous parameter in the $X$ family we have $\mu^2 \neq 2/3$ and
 $a(\mu)=2+\mu^2$, $c(\mu)=-8+4\mu^2-(102\mu^4+5\mu^6)/9 $, $d(\mu)=-8/9(2-3\mu^2)^2$.}
\label{table1}
\end{table}

The families ${\mathscr D}_{4}$ 
and ${\mathscr D}_{3,1}$ contain separable and $W$ states, respectively. 
All polynomial
invariants vanish on those states, in analogy  to the 
three-qubit case. 
For the representatives of the one-class families ${\mathscr D}_{2,2}$ and ${\mathscr D}_{2,1,1}$ 
all polynomials have identical values.  That is, they cannot
be distinguished by invariant polynomials alone although they are not SLOCC 
interconvertible. 
For the states in these families the 
invariant $\mathcal C$ depends on $\mathcal A$. According to 
equation~\eqref{eq: zero for a class} we can define a polynomial that vanishes
for these families:
\begin{equation}\label{D-invariant}
{\cal D} \ =\ {\mathcal C} + \frac{5}{9}\ {\mathcal A}^3\ \ .
\end{equation}
The family ${\mathscr D}_{1,1,1,1}$ 
contains a continuous parameter and thus infinitely many
classes. 
It can be seen from Table~I
that ${\mathcal D}(\psi_S^{(4)})\neq 0$ 
if and only if  $\psi_S^{(4)}\in {\mathscr D}_{1,1,1,1}$. 
We term this  $X$ type of entanglement after the $X$ state~\cite{OS2005,OS2010}
\begin{equation}\label{Xstate}
         \ket{\mathrm{X_4}}\ =\ \ket{0001}+\ket{0010}+\ket{0100}+\ket{1000}
                                          +\sqrt{2}\ket{1111}
 \ \ .
\end{equation}
Consequently, for symmetric four-qubit states there is a 
hierarchy of SLOCC families which can be labeled by a ``pattern'' 
$({\mathcal A}(\psi_S^{(4)}), {\mathcal D}(\psi_S^{(4)}))$ that is obtained 
from two polynomial invariants. It is tempting to call these invariants
``tangles''. There are three levels in the hierarchy: 
$({\mathcal A}=0, {\mathcal D}=0)$, 
$({\mathcal A}\neq 0, {\mathcal D}=0)$, 
$({\mathcal A}, {\mathcal D}\neq 0)$. 

\begin{table*}[htb]\label{table3}
\footnotesize
\begin{tabular}{ccccccc}
\hline \hline
 LLSS family & representative & $\mathcal A$  & ${\mathcal B}^{I}$  & ${\mathcal B}^{II}$ 
                          & $\ \  {\mathcal B}^{III}\ \ \ $  & $\mathcal C$       \\
\hline\hline
 ${\mathcal W}_{000,0_k\Psi}$\ b)   & $\ket{0000}+\ket{1101}+\ket{1110}$ 
                          &   $0$           &     $0$         &    $0$    
                          &     $0$         &        $0$                     \\
 ${\mathcal W}_{000,W}$   & $\ket{0001}+\ket{0010}+\ket{0100}+\ket{1000}$ 
                          &   $0$           &     $0$         &    $0$
                          &     $0$         &        $0$                    \\ \hline  
 ${\mathcal W}_{000,000}$   & $\ket{0000}+\ket{1111}$
                          &   $2$           &     $4$         &    $4$
                          &     $4$         &       $ -8 $                     \\ 
 ${\mathcal W}_{000,0_k\Psi}$\ a)   & $\ket{0000}+\ket{1100}+\ket{1111}$
                          &   $2$           &     $4$         &    $4$       
                          &     $4$         &       $ -8$                      \\ 
 ${\mathcal W}_{000,\mathrm{GHZ}}$  & $\ket{0\varphi\phi\psi}+\ket{1000}+\ket{1111}$
     &   $2(\varphi_0\phi_0\psi_0-\varphi_1\phi_1\psi_1)\equiv A_1$  &     $A_1^2$    &  $A_1^2$       
                          &     $A_1^2$         &       $ -A_1^3 $                     \\ 
 ${\mathcal W}_{0_k\Psi,0_j\Psi}\ a)$   & $\ket{0\phi 00}+\ket{0\phi 1\psi}+\ket{1000}
                                                                       +\ket{1101}$
     & $-2(\phi_0\psi_0+\phi_1\psi_1)\equiv A_2$           &     $A_2^2$         &    $A_2^2$
                          &     $A_2^2$         &        $-A_2^3$                      \\ 
 ${\mathcal W}_{0_k\Psi,0_j\Psi}\ b)$   & $\ket{0\phi 0\psi}+\ket{0\phi 10}+\ket{1000}
                                                                       +\ket{1101}$
                  & $-2\phi_0\equiv A_3$            &     $A_3^2$         &    $A_3^2$
                          &     $A_3^2$         &        $-A_3^3$                      \\ \hline
 ${\mathcal W}_{0_k\Psi,0_k\Psi}\ a)$   & $\ket{0000}+\ket{1100}+\lambda_1\ket{0011}
                                                                +\lambda_2\ket{1111}$
    &  $2(\lambda_1+\lambda_2)\equiv A_4$           &     $B_4^{I}$         &   $B_4^{II}\neq B_4^{I}$
             &    $B_4^{II} $        &       $-A_4^2 B_4^{II}$   \\ 
 ${\mathcal W}_{0_k\Psi,0_k\Psi}\ b)$   & $\ket{0000}+\ket{1100}+
                                          \lambda_1(\ket{0001}+\ket{0010})+$\\
                                        & $+\lambda_2(\ket{1101}+\ket{1110})$
     & $-4\lambda_1\lambda_2\equiv A_5$           &     $3 A_5^2$         &   $0$
         &    $0$        &       $0$   \\ 
 ${\mathcal W}_{0_k\Psi,\mathrm{GHZ}}$   & $\ket{0\varphi}\otimes
                                    (\ket{\phi\psi}+\ket{\bar{\phi}\bar{\psi}})+
                                     \ket{1000} +\ket{1111}$
     & $A_6$     &     $B_6^{I}$         &  $B_6^{II}\neq B_6^{I}$
  &     $B_6^{II}$      &       $-A_6^2 B_6^{II} $   \\ 
\hline
 ${\mathcal W}_{\mathrm{GHZ},W}$  & $\ket{0001}\!\!+\!\!\ket{0010}\!\!+\!\!\ket{0100}\!\!+\!\!
     \ket{1\varphi\phi\psi}\!\!+\!\!\ket{1\bar{\varphi}\bar{\phi}\bar{\psi}}$
 & $A_7$           &     $B_7^{I}$         &   $B_7^{II}$
         &    $B_7^{III}$        &       $C_7$   
\\
\hline\hline
\end{tabular}
\protect\caption{Tangle patterns for the representatives of the SLOCC 
         classification of Lamata {\em et al.}
         (cf.\ Table~I in Ref.~\cite{Lamata2007}).
         Here, $\ket{\varphi},\ldots$
         are single qubit vectors with components $(\varphi_0,\varphi_1),\ldots$
         The vectors $\ket{\varphi}$ and $\ket{\bar{\varphi}},\ldots$ are 
         linearly independent.
         Note that the reprensentative
         in line 8 coincides with a cluster state for $\lambda_1=1,\lambda_2=-1$.
         The $X$ state is an element of the family in the last line.
         For brevity, the explicit expressions 
          for $A_6,A_7, B_4^{I},B_4^{II},\ldots,B_7^{III}, C_7$ are 
         omitted.
Remarkably, 
we obtain precise functional dependences between the polynomials 
for many of the LLSS families. 
}
\end{table*}
\normalsize
{\em Polynomials and general four-qubit states. -- }
This result raises the question whether 
the polynomial classification scheme can be extended beyond symmetric states. 
Therefore, we inspect the SLOCC families in the classification due to Lamata {\em
et al.}~\cite{Lamata2007} (LLSS). 
In Table II we have listed all eleven representatives
and corresponding tangle patterns for the eight LLSS families.
The polynomials reproduce the LLSS scheme almost identically.
The functional dependences
suggest a new grouping of the states according to the tangle pattern
using the invariants~\cite{Luque2003}
\begin{eqnarray}\label{LMN}
{\mathcal L}& = &\frac{1}{48}({\mathcal B}^{II}-{\mathcal B}^{III}) \ \ \ \ \
        {\mathcal M} = \frac{1}{48}({\mathcal B}^{III}-{\mathcal B}^{I}) \nonumber \\
{\mathcal N}& = &\frac{1}{48}({\mathcal B}^{I}-{\mathcal B}^{II}) \\
%\
{\mathcal X}& = & ({\mathcal C}+{\mathcal A}^{3})^2
                          -128{\mathcal A}^2({\mathcal L}^2+{\mathcal M}^2+{\mathcal N}^2)
                                                       \nonumber \ \ ,
\end{eqnarray}
which remove the redundant functional dependences. 
With these invariants we {\em define} a hierarchical ordering into families according
to the tangle pattern displayed in Table III. This is our central result.
\begin{table}[htb]
\begin{tabular}{c||c|ccc|c}
\hline \hline
  type & $\mathcal A$ & \ \ \ $\mathcal L$\ \ \   &\ \ \  $\mathcal M$\ \ \  &
                        \ \ \  $\mathcal N$\ \ \  & $\ \ \ \mathcal X$\ \ \  \\
\hline
 $W$  &    $  0$ &   $0$   &  $0$       &    $0$    & $0$    \\
 GHZ  &    $A\neq 0$   &   $0$   &  $0$       &    $0$    & $0$  \\
 cluster  &    $   A$ &     $L\neq 0$\ \ {\rm or}  & 
                            $M\neq 0$   &    $-L-M$    & $0$ \\
 $X$  &    $  A$ &  $L$  &  $M$   &          $-L-M$    &    $X \neq 0 $\\
\hline\hline
\end{tabular}
\caption{The four-qubit SLOCC families defined via the tangle patterns
         of the invariants in equation~\eqref{LMN}.
         The invariant of highest non-vanishing degree determines the family
         to which a state belongs. We have named its
         entanglement type after a well-known state in the family.
        }
\label{table4}
\end{table}
Notice the apparent analogy between this hierarchy and the one 
for the symmetric states using the 
invariants $\mathcal A$ and $\mathcal D$, although the 
corresponding families certainly differ. 
{\em Discussion. -- }
The analysis of the tangle patterns for two different 
SLOCC classifications has lead us to a new SLOCC classification
of four-qubit states based on polynomial invariants. 
It represents an independent classification
method in its own right with several evident advantages. 

{\em (i)} In contrast to all known SLOCC classifications it is 
straightforward to decide to which family a given arbitrary state belongs.

{\em (ii)} The tangle patterns characterize  types of entanglement.
Most strikingly,
they provide not just a qualitative, but even a
quantitative description. According to Ref.~\cite{Verstraete2003},
the modulus of
any degree-2 invariant is
an entanglement monotone~\cite{Vedraletal1997,Vidal2000}. That is, by
choosing the
absolute value of the
appropriate power
for each polynomial, the tangles of the pattern 
characterize quantitatively the types of multipartite entanglement
contained in a given state.

{\em (iii)} 
Note that {\em any} (even incomplete)
set of independent polynomials~\cite{OS2005,Dokovic2009} 
generates a corresponding classification.
Our scheme displays a
 flexibility towards 
 distinguishing certain
desired  properties:  
By comparing the classifications considered above one sees
that an appropriate choice of polynomials can emphasize 
certain properties of the states in the families. 
It is particularly interesting that
the polynomials of lowest degrees 2 and 4 
separate peculiar states like GHZ and cluster states.

{\em (iv)} All these considerations can be extended to 
arbitrary multipartite systems with finite local dimensions.

{\em Acknowledgements. -- }
This work was supported by the German Research Foundation within SFB 631 (CE),
the German Academic Exchange Service (OV), and Basque Government grant
IT-472.
The authors thank A.\ Osterloh for continued  stimulating discussions,
L.\ Lamata for helpful comments, and 
K.\ Richter and J.\ Fabian 
for their support of this research.
OV thanks the QUINST group in Bilbao for their hospitality.


\begin{thebibliography}{99}
%
\bibitem{Duer2000} 
   W.\ D\"ur, G.\ Vidal, and J.I.\ Cirac,
   Phys.\ Rev.\ A {\bf 62}, 062314 (2000).
%
\bibitem{Sudbery1999} 
   H.A.\ Carteret, N.\ Linden, S.\ Popescu, and A.\ Sudbery, 
   Foundations of Physics, {\bf 29}, 527 (1999).
%
\bibitem{Verstraete2002} F.\ Verstraete, J.\ Dehaene, B.D.\ Moor, and H.\ Verschelde, 
   Phys.\ Rev.\ A \textbf{65}, 052112 (2002). 
%
\bibitem{Miyake2004} A.\ Miyake and F.\ Verstraete, 
   Phys.\ Rev.\ A \textbf{69}, 012101 (2004). 
%
\bibitem{OS2005} A.\ Osterloh and J.\ Siewert, 
   Phys.\ Rev.\ A \textbf{72}, 012337 (2005).
%
\bibitem{Lamata2006} L.\ Lamata, J.\ Le\'{o}n, D.\ Salgado, and E.\ Solano, 
   Phys.\ Rev.\ A \textbf{73}, 052322 (2006). 
%
\bibitem{Mandilara2006}
   A.\ Mandilara, V.A.\ Akulin, A.V.\ Smilga, and L.\ Viola,
   Phys.\ Rev.\ A {\bf 74}, 022331 (2006).
%
%
\bibitem{Lamata2007} L.\ Lamata, J.\ Le\'{o}n, D.\ Salgado, and E.\ Solano, 
   Phys.\ Rev.\ A \textbf{75}, 022318 (2007).
%
\bibitem{Cao2007} 
   Y.\ Cao and A.M.\ Wang,
   Eur.\ Phys.\ J.\ D\ \textbf{44}, 159 (2007).
%
\bibitem{Chterental2007}
   O.\ Chterental and D.\v{Z}. \DJo,
   in {\em Linear Algebra Research Advances}, edited by G.D.\ Ling
   (Nova Science Publishers, Inc., Hauppauge, NY, 2007), Chap.\ 4,
   133-167.
%
\bibitem{Bastin2009} T.\ Bastin, S.\ Krins, P.\ Mathonet, M.\ Godefroid, L.\ Lamata, E.\ Solano, 
   Phys.\ Rev.\ Lett.\ \textbf{103}, 070503 (2009).
%
\bibitem{Duff2010} 
   L.\ Borsten, D.\ Dahanayake, M.J.\ Duff, A.\ Marrani, and W.\ Rubens,
   Phys.\ Rev.\ Lett.\ \textbf{105}, (2010).
%
\bibitem{Li2009} D.\ Li, X.\ Li, H.\ Huang, and X.\ Li, 
   Quant.\ Inf.\ Comp.\ \textbf{9}, 0778 (2009).
%
\bibitem{Kraus2010} B.\ Kraus,
   Phys.\ Rev.\ Lett.\ {\bf 104}, 020504 (2010).
%
\bibitem{Winter2010} 
   L.\ Chen, E.\ Chitambar, R.\ Duan, Z.\ Ji, and A.\ Winter,
   Phys.\ Rev.\ Lett.\ \textbf{105}, 200501 (2010).
%
\bibitem{Luque2003} 
   J.-G.\ Luque and J.-Y.\ Thibon, 
   Phys.\ Rev.\ A \textbf{67}, 042303 (2003).
%
\bibitem{Verstraete2003} 
   F.\ Verstraete, J.\ Dehaene, and B.\ De Moor, 
   Phys.\ Rev.\ A \textbf{68}, 012103 (2003).
%
\bibitem{Wootters1997} 
   S.\ Hill and W.K.\ Wootters, 
   Phys.\ Rev.\ Lett.\ \textbf{78}, 5022 (1997).
%
\bibitem{CKW2000} 
   V.\ Coffman, J.\ Kundu, and  W.K.\ Wootters, 
   Phys.\ Rev.\ A \textbf{61}, 052306 (2000).
%
\bibitem{Wong2001} 
   A.\ Wong and N.\ Christensen, 
   Phys.\  Rev.\  A \textbf{63}, 044301 (2001).
%
\bibitem{Leifer2004} 
   M.S.\ Leifer, N.\ Linden, and A.\ Winter,
   Phys.\ Rev.\ A \textbf{69}, 052304 (2004).
%
\bibitem{Gour2010} 
   G.\ Gour,
   Phys.\ Rev.\ Lett.\ \textbf{105}, 190504 (2010).
%
\bibitem{Brylinski2002}
   J.-L.\ Brylinski and R.\ Brylinski,
   {\em Mathematics of Quantum Computation} (Chapman \& Hall, London/CRC,
            Boca Raton, FL, 2002), Chap.\ 11.
%
\bibitem{EisertBriegel2001} 
   J.\  Eisert and H.\ Briegel, 
   Phys.\ Rev.\ A \textbf{64}, 022306 (2001).
%
\bibitem{Dokovic2009} 
   D. \v{Z}. \DJo \  and A.\ Osterloh, 
   J.\  Math.\ Phys.\ \textbf{50}, 033509 (2009). 
%
%
\bibitem{OS2010}
   A.\ Osterloh and J.\ Siewert, 
   New J.\ Phys.\ \textbf{12}, 075025 (2010).
%
%
%
\bibitem{Vedraletal1997} 
   V.\ Vedral, M.B.\ Plenio, M.A.\ Rippin and P.L.\ Knight, 
   Phys.\ Rev.\ Lett.\ \textbf{78}, 2275 (1997). 
%
\bibitem{Vidal2000} 
   G. Vidal, 
   J.\ Mod.\ Opt.\ \textbf{47}, 355 (2000).
%
\end{thebibliography}
\end{document}